\documentclass[twocolumn,showpacs,preprintnumbers,amsfonts,amsmath,amssymb]{revtex4}
\usepackage{graphics}
\usepackage{epsfig}
\def\ut.#1{\mathop{#1}\limits_{\raisebox{0.4ex}[0pt]{$\sim$}}}

\begin{document}
\title{Bose--Einstein condensate of kicked rotators with time-dependent interaction}
\author{B. Mieck}
\email{mieck@theo-phys.uni-essen.de}
\author{R. Graham}
\email{graham@uni-essen.de}
\affiliation{Department of Physics in Essen,
             University Duisburg--Essen, Universit{\"a}tsstrasse 5,45117 Essen, Germany}

\begin{abstract}
A modification of the quantum kicked rotator is suggested with a time-dependent delta-kicked interaction parameter
which can be realized by a pulsed turn-on of a Feshbach resonance. The mean kinetic energy increases exponentially
with time in contrast to a merely diffusive or linear growth for the first few kicks for the quantum kicked 
rotator with a constant interaction parameter. A recursive relation is derived in a self-consistent
random phase approximation which describes this superdiffusive growth of the kinetic energy and is compared 
with numerical simulations. Unlike in the case of the quantum rotator with constant interaction, a Lax pair 
is not found. In general the delta-kicked interaction is found to lead to strong chaotic behaviour.
\end{abstract}
\maketitle

%

The realization of Bose-Einstein condensation (BEC) of dilute gases has opened new opportunities for studying
dynamical systems in the presence of many-body interactions. Previous investigations have addressed the effect
of the nonlinearities due to the interactions on the dynamical localization and on the integrability of the
one-dimensional Gross-Pitaevskii equation (GP-equation) in the presence of a time-periodic delta-kicked
external potential $v(q,t)$ \cite{R1,G1}. As is well-known 
the classical counterpart of the quantum kicked rotator without interaction
displays chaotic motion, leading to diffusive growth in the kinetic energy above a certain value of the 
kick-strength $K$ of the external potential \cite{C1,C2}. In the corresponding quantum kicked rotator 
after a few initial kicks, during which the mean kinetic energy increases roughly linearly, the motion 
becomes quasiperiodic (barring special cases where quantum resonances occur) with a dynamical localization of the 
kinetic energy \cite{N1}-\cite{N6}. 
The infinite number of conserved quantities, leading to the integrability of the quantum
kicked rotator, are the probabilities with which the quasi-energy states are contained in a given initial state.
The inclusion of the nonlinear GP-term to the Schr\"{o}dinger equation in numerical simulations 
was found to give rise to chaotic behaviour and delocalization of the kinetic energy \cite{R1,G1}. However, 
analytical considerations still show the existence of an infinite number of independent integrals of motion 
for the one-dimensional GP-equation with an arbitrary external potential \cite{T1,G1}.
At the moment it seems difficult to reconcile this analytical result with the results of the numerical
simulations. 

In this paper we suggest and investigate a modification of the quantum kicked rotator by introducing a
time-dependent interaction $g(t)$ of delta-kicks. Furthermore, we conjecture that the numerically observed
delocalization of the quantum kicked rotator with constant interaction parameter $g$ could be a consequence
of employing discretizations of the nonlinear Schr\"{o}dinger equation which do not preserve the property of
integrability of the GP-equation with external potential in the continuum.

The scaled GP-equation with external potential 
\(v(q,t)=K\cos(q)\sum_{n=-\infty}^{n=\infty}\delta(t-n)\)
\begin{equation} \label{1}
\imath\hbar\dot{\psi}=-\frac{\hbar^{2}}{2}\nabla_{q}^{2}\psi+v(q,t)\;\psi+g\;|\psi|^{2}\;\psi
\end{equation}
is modified by introducing a time-dependent interaction parameter $g(t)$
\begin{equation} \label{2}
\imath\hbar\dot{\psi}=-\frac{\hbar^{2}}{2}\nabla_{q}^{2}\psi+v(q,t)\;\psi+g(t)\;|\psi|^{2}\;\psi.
\end{equation}
The time-dependent interaction can be achieved by a suitable time-dependent tuning to a Feshbach resonance \cite{O1}.
By a short pulsed modulation of the magnetic field used for the tuning, it is possible to taylor a time-dependent
coupling parameter of the form \(g(t)=g\sum_{n=-\infty}^{n=\infty}\delta(t-n)\) which we want to analyze here. Thus the BE-condensate,
we consider in the following, receives additional delta 
kicks apart from that of the external potential already present for the quantum kicked rotator. In between two kicks, 
the BE-condensate, described by (\ref{2}) follows free motion with the quantum kinetic energy, whereas in the original
quantum kicked rotator with constant $g$ (Eq. (\ref{1})), the motion is determined by the integrable nonlinear
Schr\"{o}dinger equation (NLS) without an external potential. Therefore, numerical simulations become
considerably more efficient and easier for Eq. (\ref{2}) with delta-kicked interaction than for Eq. (\ref{1})
because the latter equation with permanent nonlinearity needs many integration steps between two kicks \cite{R1,G1}
\footnote{We are grateful to L. Santos, who suggested this idea to us in a private discussion.}.

The map for the wavefunction, stroboscopically taken after each kick $t_{n}$, with delta interaction $g(t)$
is given by
\begin{eqnarray} \label{3}
\psi(q,t_{n+1}^{+}) &=& e^{-\imath/\hbar\cdot V(q,t_{n+1}^{-})}\;e^{\imath\hbar/2\cdot\nabla_{q}^{2}}\;
\psi(q,t_{n}^{+}) \\ \label{4}
V(q,t_{n+1}^{-}) &=& K\;\cos(q)+g\;|\psi(q,t_{n+1}^{-})|^{2}\; ,
\end{eqnarray}
where the $t_{n}^{+}$, $t_{n}^{-}$ variables refer to times immediately after (+) and before (-) the
occurrence of the n-th kick \footnote{We assume here that the two types of kicks (of the potential and of the interaction)
occur at the same time, but this is of no importance for our conclusions.}
\begin{equation} \label{5}
\psi(q,t_{n+1}^{+})=e^{-\imath/\hbar\cdot V(q,t_{n+1}^{-})}\;\psi(q,t_{n+1}^{-})\; .
\end{equation}

Formally, Eq. (\ref{2}) with a general interaction parameter $g(t)$ can be transformed into Eq. (\ref{1})
by substitution of the wavefunction \(\sqrt{g(t)/g_{0}}\;\psi(q,t)=\widetilde{\psi}(q,t)\). In terms of the
new wavefunction \(\widetilde{\psi}(q,t)\), the GP-equation acquires a spatially constant, but imaginary
time-dependent potential part
\begin{eqnarray} \label{6}
\imath\hbar\dot{\widetilde{\psi}} &=&-\frac{\hbar^{2}}{2}\nabla_{q}^{2}\widetilde{\psi}+\widetilde{v}(q,t)\;\widetilde{\psi}+
g_{0}\;|\widetilde{\psi}|^{2}\;\widetilde{\psi} \\ \label{7} 
\widetilde{v}(q,t) &=& v(q,t)+\imath\frac{\hbar}{2}\frac{\dot{g}(t)}{g(t)}\; .
\end{eqnarray}
However, a Lax pair for the GP-equation with a {\it complex} potential $\widetilde{v}(q,t)$
and constant interaction $g_{0}$ cannot be constructed as in \cite{T1,G1} for a {\it real} 
external potential $v(q,t)$:

Using the following general ansatz with \(2\times 2\) matrices for the generator $\mathcal{X}(q,t)$ and
$\mathcal{T}(q,t)$ of a Lax-pair
\begin{eqnarray} \label{8}
\mathcal{X}(q,t) &=&\left(
\begin{array}{cc}
-\imath\;k_{1}(q,t) & \sqrt{g_{0}}\;\widetilde{\psi}^{*}(q,t) \\
\sqrt{g_{0}}\;\widetilde{\psi}(q,t) & \imath\;k_{2}(q,t)
\end{array}\right) \\ \label{9}
\mathcal{T}(q,t) &=&\left(
\begin{array}{cc}
A(q,t) & B(q,t) \\
C(q,t) & D(q,t)
\end{array}\right)_{,}
\end{eqnarray}
with yet undetermined complex functions $A$, $B$, $C$, $D$ and $k_{1}$, $k_{2}$,
the compatibility of the pair of equations
\begin{equation} \label{10} 
\hbar\frac{\partial w}{\partial q}=\mathcal{X}\;w\hspace*{0.75cm}\hbar\frac{\partial w}{\partial t}=\mathcal{T}\;w
\end{equation}
requires the following equations to be satisfied
\begin{eqnarray} \label{11} \hspace*{-0.9cm}
\imath\hbar\dot{\widetilde{\psi}} &=&\imath\frac{\hbar}{\sqrt{g_{0}}}C_{q}+\imath(D-A)\;\widetilde{\psi}+\frac{(k_{1}+k_{2})}{\sqrt{g_{0}}}\;C \\
\label{12} -\imath\hbar\dot{\widetilde{\psi}}^{*} &=&-\imath\frac{\hbar}{\sqrt{g_{0}}}B_{q}+\imath(D-A)\widetilde{\psi}^{*}+
\frac{(k_{1}+k_{2})}{\sqrt{g_{0}}}\;B \\ \label{13} \imath\hbar \dot{k}_{1} &=& -\hbar\;A_{q} + \sqrt{g_{0}}
(C\;\widetilde{\psi}^{*}-B\;\widetilde{\psi}) \\ \label{14} \imath\hbar\dot{k}_{2} &=& \hbar\;D_{q}+\sqrt{g_{0}}
(C\;\widetilde{\psi}^{*}-B\;\widetilde{\psi}).
\end{eqnarray}
Comparing Eqs. (\ref{11}) and (\ref{12}) to (\ref{6},\ref{7}) and its complex conjugate, the terms
\(\imath(D-A)\widetilde{\psi}\) and \(\imath(D-A)\widetilde{\psi}^{*}\) in (\ref{11},\ref{12}) have to be identified with
\(\widetilde{v}(q,t)\;\widetilde{\psi}\) and \(\widetilde{v}^{*}(q,t)\;\widetilde{\psi}^{*}\) and do not allow
for an imaginary part of \(\widetilde{v}(q,t)\). Due to the missing of a Lax-pair, complete chaotic
behaviour of the GP-equation with a general time-dependent interaction $g(t)$ can be expected. This is
confirmed by our numerical simulations.

It is possible to derive an approximate recursion relation for the mean kinetic energy, just taken
after kick time $t_{n}^{+}$
\begin{equation} \label{15}
\langle p^{2}(t_{n}^{+})\rangle = \int_{0}^{2\pi}dq\;|\partial_{q}\psi(q,t_{n}^{+})|^{2}
\end{equation}
and time $t_{n+1}^{+}$, \(\langle p^{2}(t_{n+1}^{+})\rangle\). The derivation is mainly based on the assumption that
kick-to-kick correlations between $\psi(q,t_{n+1}^{+})$ and $\psi(q,t_{n}^{+})$ can be neglected. Applying the
expression $V(q,t_{n+1}^{-})$ for the external potential and interaction before the (n+1)-th kick (\ref{3},\ref{4}), the
averaged kinetic energy $\langle p^{2}(t_{n+1}^{+})\rangle$ can be related to the wavefunction of the previous kick
\begin{eqnarray} \label{16}
\partial_{q}\psi(q,t_{n+1}^{+}) &=&-\frac{\imath}{\hbar}\partial_{q}V(q,t_{n+1}^{-})\;\psi(q,t_{n+1}^{+}) \\ \nonumber &+&
e^{-\imath/ \hbar\cdot V(q,t_{n+1}^{-})}\;e^{\imath\hbar/2\cdot\nabla_{q}^{2}}\;\partial_{q}\psi(q,t_{n}^{+})
\end{eqnarray}
\begin{eqnarray} \nonumber
\lefteqn{\langle p^{2}(t_{n+1}^{+})\rangle =\frac{1}{\hbar^{2}}\int_{0}^{2\pi}dq\;(\partial_{q}V(q,t_{n+1}^{-}))^{2}\;
|\psi(q,t_{n+1}^{+})|^{2} } \\ \nonumber &+& \int_{0}^{2\pi}dq\;
\Big|e^{\imath\hbar/2\cdot\nabla_{q}^{2}}\;\partial_{q}\psi(q,t_{n}^{+})\Big|^{2} 
\\ \label{17} &-&\frac{2}{\hbar}
\int_{0}^{2\pi}dq\;\Im\bigg[\partial_{q}V(q,t_{n+1}^{-})\;\psi^{*}(q,t_{n+1}^{+}) \\ \nonumber &\times&
\Big(e^{-\imath/ \hbar\cdot V(q,t_{n+1}^{-})}\;e^{\imath\hbar/2\cdot\nabla_{q}^{2}}\;\partial_{q}\psi(q,t_{n}^{+})\Big)\bigg].
\end{eqnarray}
The last integral in (\ref{17}) contains only a single wavefunction \(\psi^{*}(q,t_{n+1}^{+})\) at time $t_{n+1}$
so that, due to the assumption of independent phases between neighbouring kicks, this term approximately
vanishes. In the second term of relation (\ref{17}), the spatial derivative \(\partial_{q}\psi(q,t_{n}^{+})\) is
transformed by a unitary operator which cancels in the integral so that the integrand becomes
\(|\partial_{q}\psi(q,t_{n}^{+})|^{2}\), and in consequence the second term is \(\langle p^{2}(t_{n}^{+})\rangle\). 
Taking the spatial derivative \(\partial_{q}V(q,t_{n+1}^{-})\) in the first term of
\(\langle p^{2}(t_{n+1}^{+})\rangle\) (\ref{17})
\begin{eqnarray} \nonumber
\partial_{q} V(q,t_{n+1}^{-}) &=& -K\;\sin(q)+g\;\partial_{q}\psi^{*}(q,t_{n+1}^{-})\;\psi(q,t_{n+1}^{-}) \\ \label{18} &+&
g\;\psi^{*}(q,t_{n+1}^{-})\;\partial_{q}\psi(q,t_{n+1}^{-}),
\end{eqnarray}
we only keep expressions with absolute values of wavefunctions and their derivatives and obtain
\begin{eqnarray} \label{19}
\lefteqn{\hbar^{2}\;\langle p^{2}(t_{n+1}^{+})\rangle \approx  K^{2}\int_{0}^{2\pi}dq\;\sin^{2}(q)\;
\underbrace{|\psi(q,t_{n+1}^{+})|^{2}}_{1/(2\pi)}} \\
\nonumber &+& 2\;g^{2}\int_{0}^{2\pi}\hspace*{-0.3cm}dq\;
|\partial_{q}\psi(q,t_{n+1}^{-})|^{2}\;\underbrace{|\psi(q,t_{n+1}^{-})|^{2}}_{1/(2\pi)}\;
\underbrace{|\psi(q,t_{n+1}^{+})|^{2}}_{1/(2\pi)} \\ \nonumber &+& \hbar^{2}\;
\langle p^{2}(t_{n}^{+})\rangle ,
\end{eqnarray}
where the absolute values of $\psi(q,t_{n+1}^{-})$ and $\psi(q,t_{n+1}^{+})$ are replaced by the mean density
\(1/(2\pi)\). 
Since the wavefunction \(\psi(q,t_{n+1}^{-})\) differs from \(\psi(q,t_{n}^{+})\) only by the
time development with \(\exp\{\imath\hbar/2\cdot\nabla_{q}^{2}\}\), the average of \( |\partial_{q}\psi(q,t_{n+1}^{-})|^{2}\) can
be replaced by \( |\partial_{q}\psi(q,t_{n}^{+})|^{2}\).

This results in a recursive relation between \(\langle p^{2}(t_{n+1}^{+})\rangle\) and
\(\langle p^{2}(t_{n}^{+})\rangle\)
\begin{equation} \label{20}
\langle p^{2}(t_{n+1}^{+})\rangle = \frac{K^{2}}{2\hbar^{2}}+\left(1+\frac{g^{2}}{2\pi^{2}\hbar^{2}}\right)\;\langle p^{2}(t_{n}^{+})\rangle .
\end{equation}
In the case of vanishing interaction parameter $g$, \(\langle p^{2}(t_{n}^{+})\rangle\) increases linearly with
\(K^{2}/(2\hbar^{2})\), as expected for the initial classical diffusive regime of the original quantum
kicked rotator. However, the phenomenon of quantum localization cannot be derived from the recursive
relation which assumes no kick-to-kick correlation of the phases of the wavefunctions at time $t_{n}$ and
$t_{n+1}$ and neglects the correlations due to the presence of quasi-energy eigenstates.
For finite interaction $g$ quasi-energy states no longer exist and the assumption leading to (\ref{20}) can be
consistent. Indeed, due to Eq. (\ref{20}), the mean kinetic energy follows an exponential or superdiffusive
growth, indicating strong chaotic behaviour, which makes the assumption of statistical independence
of subsequent kicks self-consistent. The continuum limit of the recursion relation (\ref{20})
also yields an exponential growth of \(\langle p^{2}(t)\rangle\)
\begin{equation} \label{21}
\langle p^{2}(t)\rangle = \frac{\pi^{2}K^{2}}{g^{2}}\left[\exp\left(\frac{g^{2}}{2\hbar^{2}\pi^{2}}\;t\right)-1\right]_{,}
\end{equation}
where \(\langle p^{2}(t=0)\rangle\) vanishes if the initial wavefunction \(\psi(q,t=0)=1/\sqrt{2\pi}\) is taken
with zero momentum as in our simulations. Relation (\ref{21}) for \(\langle p^{2}(t)\rangle\) interpolates between purely
linear increase for small $g$ and exponential growth for strong interactions.

\begin{figure}[b]
\includegraphics[width=8.0cm, height=5.0cm]{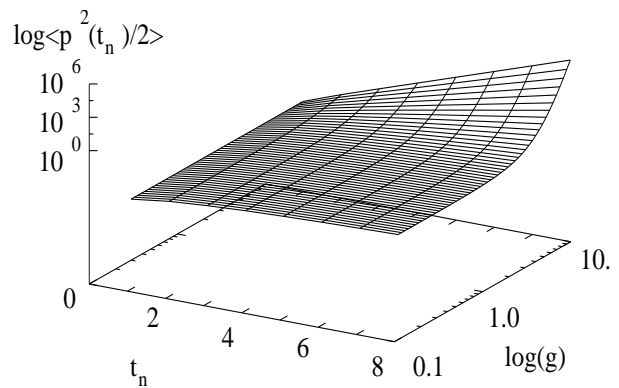}
\caption{\label{f1} Logarithm of the mean kinetic energy determined from the recursive relation (\ref{20}) for vanishing
initial momentum of the wavefunction \(\psi(q,t_{0})=1/\sqrt{2\pi}\) with a kick strength of the external potential of
\(K=1.0\). \(\log\langle p^{2}(t_{n})/2\rangle\) is displayed for the first eight kicks $t_{n}$ versus the logarithm of
the parameter $g$ of the delta-kicked interaction, ranging from \(g=0.1\) to \(g=10.0\).}
\end{figure}

\begin{figure}
\includegraphics[width=8.0cm,height=5.0cm]{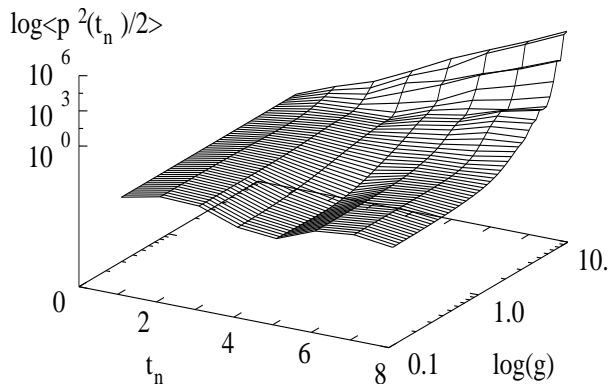}
\caption{\label{f2} Results of the numerically obtained averaged kinetic energy on a logarithmic scale for the same parameters
as in Fig. \ref{f1}. (Configurations of the axes are the same as in Fig. \ref{f1}.) 
Qualitative agreement for the superdiffusive growth of \(\langle p^{2}(t_{n})/2\rangle\)
is obtained with the exponential relation (\ref{20}).}
\end{figure}

\begin{figure}
\includegraphics[width=8.0cm,height=5.0cm]{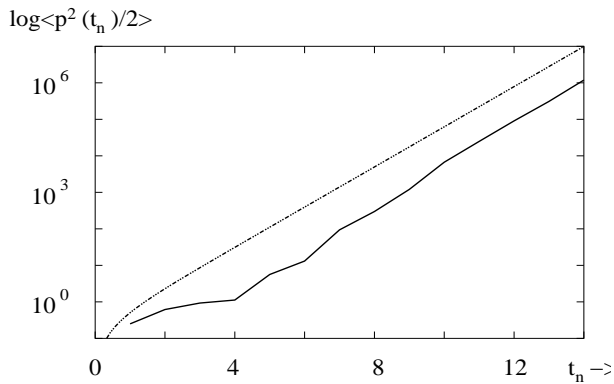}
\vspace*{0.1cm}
\caption{\label{f3} Comparison of the exponential relation (\ref{21}) (dash-dotted upper line) 
for \(g=5.0\) of the delta-kicked interaction and the kick strength \(K=1.0\) with the corresponding 
numerical simulations (lower solid line). The slope of the exponential increase of the kinetic energy,
obtained by (\ref{21}), is in good agreement with the computations whereas absolute values differ by
a constant factor.}
\end{figure}

The comparison of the recursive relation for \(\langle p^{2}(t_{n}^{+})\rangle\) (\ref{20}) and \(\langle p^{2}(t)\rangle\)
(\ref{21}) with numerical simulations yields a qualitative agreement on a logarithmic scale (see Figs. \ref{f1} to \ref{f3}).
In Fig. \ref{f3} we display on a logarithmic scale the mean kinetic energy for the delta-kick interaction
\(g(t)=g\sum_{n=-\infty}^{n=\infty}\delta(t-n)\) with \(g=5\) and for the external potential
\(v(q,t)=K\cos(q)\sum_{n=-\infty}^{n=\infty}\delta(t-n)\) with kick strength \(K=1\). Our estimate (\ref{20}) and (\ref{21})
for \(\langle p^{2}(t_{n}^{+})\rangle\) gives a consistent value for the slope whereas the absolute values of
\(\langle p^{2}(t_{n}^{+})\rangle\) differ by a constant factor from the numerical results. 
This deviation may result from the substitution of the average of the product of the densities with the product of the average
of the wavefunctions (cf. Eq. (\ref{19})). The phenomenon of
quantum localization as for vanishing interaction $g=0$ or small but constant $g$ has not been obtained in our simulations
with available number of space points up to $2^{16}$ on the periodic interval \([0,2\pi)\).The increase of the mean kinetic
energy only ended at the maximum momentum squared which is limited by the spatial intervals.
The complete missing of any sign of quantum localization and the superdiffusive increase of 
\(\langle p^{2}(t)\rangle\) are the main differences
to the original quantum kicked rotator. In the case of constant interaction parameter \(g(t)=g\) (Eq. (\ref{1})), 
a slow increase of the mean kinetic energy can be observed in numerical simulations \cite{R1,G1}. We conjecture that
that this delocalization may be caused by a nonintegrable discretization. 
According to Ref. \cite{T1} and our considerations,
the obvious discrete version of the GP-equation (\ref{1}) is not integrable (\(\psi_{n}(t)=\psi(q_{n},t)\))
\begin{equation} \label{22}
\imath\hbar\dot{\psi}_{n}=-\frac{\hbar^{2}}{2}\frac{\psi_{n+1}-2\psi_{n}+\psi_{n-1}}{(\Delta q)^{2}}+v_{n}(t)\;\psi_{n}+g\;|\psi_{n}|^{2}\;\psi_{n}
\end{equation}
and does not possess a Lax-pair, whereas the following discrete Schr\"{o}dinger equation 
with constant interaction parameter $g$
can be derived from a compatibility condition with appropriate generators $\mathcal{X}$ and $\mathcal{T}$ for a Lax-pair
\begin{eqnarray}\label{23}
\lefteqn{\imath\hbar\dot{\psi}_{n}=-\frac{\hbar^{2}}{2}\frac{\psi_{n+1}-2\psi_{n}+\psi_{n-1}}{(\Delta q)^{2}} }
\\ \nonumber &+&\frac{1}{2}(v_{n}(t)+v_{n+1}(t))\;\psi_{n} +\frac{g}{2}\;|\psi_{n}|^{2}\;(\psi_{n+1}+\psi_{n-1}).
\end{eqnarray}
The mean kinetic energy derived from (\ref{23}) and (\ref{22}) can therefore be expected to behave quite
differently, but (\ref{23}) is, unfortunately, more difficult to simulate accurately than (\ref{22}), and has
not been analyzed so far.

Discrete versions of differential equations need not have in general the same integrability or chaoticity properties
as their continuous counterparts \cite{T1}. However, the modification of the quantum kicked rotator by a delta-kicked 
interaction, we have studied here, 
is free from such difficulties since it gives a unique
map from the continuum equation (\ref{2}) to the time development (\ref{3}) of the wavefunction with only free motion
between two kicks. The superdiffusive growth and strong delocalization of the mean kinetic energy should be observable
experimentally by a suitable tuning to a Feshbach resonance. 
A realization of a ring structure for a BE-condensate, on which
short potential kicks are applied, has been suggested \cite{G1}. 
Qualitatively, a stronger expansion rate of the condensate should 
be observed for the delta-kicked interaction because of the exponential increase of the mean kinetic energy.


\begin{acknowledgments}
 This work was supported by the Deutsche Forschungsgemeinschaft within the
 SFB/TR 12. We gratefully acknowledge the discussion with L. Santos who suggested the use of a delta-kicked
interaction. 
\end{acknowledgments}

%
%


\begin{thebibliography}{}
\bibitem{R1}
         Ch.Zhang, J.Liu, M.G.Raizen, and Q.Niu, Phys. Rev. Lett {\bf 92}, 054101 (2004)
\bibitem{G1}
         B. Mieck and R. Graham, J. Phys. A {\bf 37} L581-L588
\bibitem{C1}
         B.V.Chirikov, Phys. Reports {\bf 52}, 265 (1979)
\bibitem{C2}
         J.M.Greene, J. Math. Phys. {\bf 20}, 1183 (1979)
\bibitem{N1}
         G. Casati, B.v.Chirikov, J.Ford, and F.M.Izrailev, in 'Lecture Notes in Physics 93', G. Casati and J.Ford eds., Springer, Berlin 1979; B.V.Chirikov, F.M.Izrailev, and D.L.Shepelyansky, Sov. Sci. Rev. {\bf 2C}, 209 (1981)
\bibitem{fis}
         D.R.Grempel, R.E.Prange, and S.Fishman, Phys. Rev. A {\bf 29}, 1639 (1984)
\bibitem{schl}
         R.Graham, M.Schlautmann, and P.Zoller, Phys. Rev. A{\bf 45}, R19 (1992) 
\bibitem{N6}
         F.L.Moore, J.C.Robinson, C.Bharucha, P.E.Williams, and M.G.Raizen, Phys. Rev. Lett. {\bf 73}, 2974 (1994); F.L.Moore, J.C.Robinson, C.F.Bharucha, B.Sundaram, and M.G.Raizen, Phys. Rev. Lett. {\bf 75}, 4598 (1995)
\bibitem{T1} M.J.Ablowitz, B.Prinari, and A.D.Trubatch, 'Discrete and Continuous Nonlinear Schroedinger Systems', London Math. Soc. Lecture Note Series 302, Cambridge University Press, Cambridge 2004
\bibitem{O1} M. Olshanii, Phys. Rev. Lett. {\bf 81}, 938 (1998)
\end{thebibliography}
\end{document}